\documentstyle[preprint,aps,prbbib]{revtex}
\tightenlines

\def\T{\begin{mathcal}T\end{mathcal}}

\def\mstar{m^*/m_B}

\def\wps2{\omega_{p}^{*\,2}}

\def\sig1w{\sigma_{1}(\omega)}

\def\johnsquare{\kern1pt\vbox{\hrule height 0.6pt\hbox{\vrule width 
0.6pt\hskip 2pt \vbox{\vskip 4pt}\hskip 2pt\vrule width 0.6pt}\hrule
height 0.6pt}\kern1pt}

\begin{document}
\title{Temperature dependent scattering rate and optical mass of
ferromagnetic metallic manganites} \author{J.~R.~Simpson, H.~D.~Drew,
V.~N.~Smolyaninova, R.~L.~Greene, M.~C.~Robson, Amlan~Biswas, and
M.~Rajeswari} \address{Department of Physics,\\ University of
Maryland, College Park, MD~20742.}  \maketitle

\begin{abstract}
  We report on the optical properties of the hole-doped manganites
  La$_{0.7}$Ca$_{0.3}$MnO$_{3}$\ and La$_{0.7}$Sr$_{0.3}$MnO$_{3}$.
  Transmission and reflection of thin films are measured in the
  infrared at temperatures from 10~-~150~K using Fourier-transform
  spectroscopy. The scattering rate and optical mass are obtained by
  fitting the far-infrared transmission data to a Drude model. The
  scattering rate shows a $T^{2}$ dependence with temperature. The
  optical mass enhancement differs only slightly from specific heat
  results. In addition, we compare the infrared spectral weight to
  band structure calculations [M. Quijada {\it et al}., Phys. Rev.  B
  {\bf 58}, 16093 (1998)].

%Draft 3 on \today\space (not for distribution).
\end{abstract}

\date{\today}

\narrowtext

%\section{Introduction}

The recent discovery of colossal magnetoresistance (CMR) in hole-doped
manganites of the form La$_{1-x}$A$_x$MnO$_3$ where La is a lanthanide
and A is an alkaline-earth element has renewed interest in this
complex magnetic system. For doping concentrations in the range
$0.2<x<0.5$, this material undergoes a phase transition from
paramagnetic insulator to ferromagnetic metal at the Curie temperature
$T_c$. The double-exchange model (DE) first proposed by
Zener\cite{zener} explains the phase transition in terms of the Mn
d-electrons, namely, the strong Hund's coupling between the three
electrons localized in the $t_{2g}$ orbitals and the 1-x electrons in
the $e_g$ orbitals.  While DE qualitatively describes the
metal-insulator transition at $T_c$, recent
experimental\cite{quijada,kaplan,billinge} and
theoretical\cite{millis} work indicates the importance of coupling
between charge and the lattice, specifically the dynamic Jahn-Teller
(JT) effect. Below $T_c$, itinerant conduction results from ferromagnetic
ordering increasing the width of the $e_g$ band and suppressing the JT
effect. Evidence for charge and orbital ordering at different doping
concentrations suggests the ground state may be the result of
competition between interactions with the lattice and different types
of ordering: ferromagnetic, charge, and orbital. At present, the exact
nature of the low-temperature state is not fully understood.

Optical conductivity
studies\cite{quijada,kaplan,nohlc,okimoto,tokura,nohns} have shown a
shift in spectral weight from the visible to the infrared as the
temperature is lowered below $T_{c}$. In the ferromagnetic state, the
low frequency optical spectrum is characterized by Drude-like
conduction.  Several groups have reported an anomalously small Drude
weight in both La$_{0.7}$Ca$_{0.3}$MnO$_{3}$ (LCMO) and
La$_{0.7}$Sr$_{0.3}$MnO$_{3}$\ (LSMO).\cite{nohlc,okimoto,tokura}
Interpreting the small Drude weight in terms of an enhanced optical
mass, the effective mass values reported in these optical studies are
much greater than the results from specific heat
measurements.\cite{smoly,smoly2} Small apparent Drude weight may also
be understood in terms of charge ordering. A charge density wave opens
a partial gap in the density of states at the Fermi level $N(E_f)$,
increasing the optical mass while decreasing the specific heat
mass. However, recent findings have cast doubts on these small Drude
weights.\cite{nohns,takenaka}

In this paper, we concentrate on the optical conductivity, scattering
rate, and mass enhancement of thin-film manganites at temperatures
below $T_{c}$.  At these temperatures, the low-frequency conductivity
exhibits a Drude-like behavior. We find an optical mass which is
comparable to the specific heat mass, a result which is inconsistent
with strong charge ordering in these optimally doped materials. The
results suggest that the metallic state of the ferromagnetic
manganites is a Fermi liquid.

%\section{Experiment}

Thin films of LCMO and LSMO are grown on LaAlO$_3$ (LAO) substrates
using pulsed laser deposition.  The LCMO film was subsequently
annealed in an O$_2$ environment.\cite{O2annealing} Low residual
resistivity $\rho(T\!=\!4~\text{K})$, and high resistivity peak
temperature $T_p$ indicate the excellent quality of the films. For
LCMO, $\rho(T\!=\!4~\text{K} )=124~\mu\Omega\,\text{cm}$ and
$T_p=275$~K, and for LSMO,
$\rho(T\!=\!4~\text{K})=15~\mu\Omega\,\text{cm}$ and $T_p>350$~K.

Transmittance $\begin{mathcal}T\end{mathcal}(\omega)$ and reflectance
$\begin{mathcal}R\end{mathcal}(\omega)$ measurements of near normal
incidence light are performed using a Fourier-transform
spectrometer.\cite{quijada} Temperature dependent spectra from
10~to~150~K ($T<T_c$) are obtained. We focus on two
frequency ranges: the far-infrared (far-IR) 2.5~-~15~meV and the
mid-infrared (mid-IR) 0.2~-~1.2~eV. The spectral gap between 15 and
200 meV is due to the opacity of LAO in this spectral
range. Determination of the film conductivity in both frequency ranges
requires knowledge of the index of refraction $n$ and extinction
coefficient of the substrate, which is measured separately.

%\section{Far-IR}

In the far-IR, the thin-film transmittance (transmission of the
film/substrate divided by transmission of the bare substrate) is given by,

\begin{equation}
\begin{mathcal}T\end{mathcal}(\omega)\approx\frac{1}{\left|1+Z(\omega)
\sigma(\omega)\right|^2} \quad ; \quad Z(\omega)=\frac{Z_0d_f}{n(\omega)+1}\ ,
\label{eq;trans}
\end{equation}

\noindent where $Z_0$ is the impedance of free space, $d_f$ is the
film thickness, and $\sigma(\omega)$ is the complex optical
conductivity. [Note that while not explicitly expressed in
Eq.~(\ref{eq;trans}), multiple reflections are included in the
analysis.]

Given the Drude-like low-frequency behavior of the conductivity below
$T_{c}$ observed in these materials\cite{quijada,nohlc,nohns}, we fit
the measured $\begin{mathcal}T\end{mathcal}$ with a simple Drude model

\begin{equation}
\sigma(\omega)=\frac{1}{4\pi}\frac{\omega_{p}^{*\,2}}{\gamma^*+i
\omega}\ ,
\label{eq;Drude}
\end{equation}

\noindent where $\gamma^*$ is the effective scattering rate and $\omega_p^*$
is the effective plasma frequency or spectral weight of the Drude
conductivity.

Transmittance curves in the far-IR (solid lines) and fits to a Drude
model (dashed lines) for several temperatures are shown in
Fig.~\ref{fig;data}. The low-frequency $\begin{mathcal}T\end{mathcal}$
increases with temperature corresponding to a decrease in the
conductivity.  At higher temperatures, $\T$ deviates less from the
limiting value (dotted lines) where $\gamma^* \rightarrow \infty$. In
the infinite scattering-rate limit,
$\sigma(\omega)\rightarrow\sigma_0$ and therefore the frequency
dependence of the limiting case is due to the $n(\omega)$ of the LAO
substrate. The slight upturn of the data relative to the fit curves
(dashed lines) above 100~cm$^{-1}$ appears independent of temperature
(Note uncertainty in $\begin{mathcal}T\end{mathcal}$ increases as the
frequency nears the substrate cutoff around 120~cm$^{-1}$).

Fitting the measured $\begin{mathcal}T\end{mathcal}$ with
Eq.~(\ref{eq;Drude}), we show the temperature dependence of the
resulting $\gamma^*$ and $ \omega_p^*$ in
Fig.~\ref{fig;fitparams}. The scattering rate has a $T^2$ temperature
dependence and is fit (solid line) with $\gamma^*(T)= \gamma^*_0+(k_B
T)^2/W$, where the fitting parameters are the defect scattering rate
at zero temperature $\gamma_0$ and the characteristic energy for
inelastic scattering $W$, and $k_B$ is Boltzmann's constant.  $W$
values for LCMO and LSMO are listed in Table~\ref{table1}. As
temperature increases, $\begin{mathcal}T\end{mathcal}$ approaches the
limiting value and the uncertainty in determining both $\gamma^*$ and
$\omega_p^*$ increases. Figure~\ref{fig;fitparams} shows error bars
for $\gamma^*$ in (a) and $\omega_p^*$ in (b). $\gamma^*$ falls below
the $T^2$ fit at the highest temperatures (140 and 150~K for LCMO and
150~K for LSMO).  At this time, we are uncertain as to the cause of
this apparent decrease in the scattering rate and whether it continues
above 150~K. While $\gamma^*$ exhibits a strong temperature
dependence, $\omega_p^*$ remains relatively independent of
temperature.

Having obtained $\gamma^*$ and $\omega_p^*$, we determine the
zero-frequency resistivity

\begin{equation}  \label{eq;resistivity}
\rho=\frac{4 \pi \, \gamma^*}{\omega_{p}^{*\,2}}=\frac{4 \pi \,
\gamma}{\omega_{p}^2}\ .
\end{equation}

\noindent Temperature dependence of $\rho$ is shown in Fig.~\ref
{fig;fitparams}(c). Both the far-IR values (solid circles) derived
from Eq.~(\ref{eq;resistivity}) and the DC values (lines) from
standard four-probe measurements are plotted for comparison. DC and
far-IR resistivity agree reasonably well in LCMO, however, the DC
value for LSMO has been scaled by a factor of 9. Similar results for
LSMO films were reported earlier\cite {quijada} and attributed to an
anisotropic conductivity resulting from substrate induced
strain. Uncertainty in the far-IR $\rho$ (error bars are nominally the
size of the points) results from uncertainty in the film thickness and
detector noise.

The optical mass enhancement is defined by

\begin{equation}  \label{eq;firmass}
m^*/m_B=(\omega_{p}^B/\omega_{p}^*)^2=1+\lambda \ ,
\end{equation}

\noindent where $\lambda$ is the mass enhancement factor and the
plasma frequencies, $\omega_{p}^B$ and $\omega_{p}^*$, are obtained
from band structure calculations and far-IR measurements,
respectively. Pickett and Singh\cite{pickett} predict
$\omega_{p}^B=1.9$~eV for LCMO. A tight binding parameterization of
the band structure with hopping parameter, $t_0=0.6$~eV, gives
$\omega_{p}^B=2.1$~eV.\cite{quijada} Taking $\omega_p^B=1.9$~eV, mass
enhancement as a function of temperature is shown in
Fig.~\ref{fig;fitparams}(d) and found to be approximately 3 for both
materials.

The $T^2$ temperature dependence of $\gamma^*$ suggests that
$\gamma^*$ may also be frequency dependent. The analysis presented
above assumed a frequency-independent $\gamma^*$ and $\omega_p^*$. A
modification to the Drude theory is necessary to consider a frequency
dependent $\gamma^*$. To obtain an optical conductivity that is a
proper response function satisfying Kramers-Kronig relations, a
frequency dependent scattering rate must include a real and imaginary
part, $\gamma(\omega)=\gamma_1(\omega)+i\gamma_2(\omega) $, where
$\gamma_2(\omega)=\omega \lambda (\omega)$. This leads to the extended
Drude model\cite{extDrude} with a renormalized scattering rate,
$\gamma^*=\gamma/(1+\lambda(\omega))$, and a renormalized plasma
frequency, $\omega_p^*=\omega_{p}/\sqrt{1+\lambda(\omega)}$. Thus, the
frequency-dependent scattering gives rise to a concomitant
frequency-dependent mass enhancement,
$m^*/m_B(\omega)=1+\lambda(\omega)$.

In order to obtain $m^*/m_B(\omega)$ for these transmission
measurements, we must assume (since we cannot measure directly) a form
for the frequency dependence of $\gamma$. The $T^2$ dependence of
$\gamma^*$ implies a similar $\omega^2$ dependence at
low-frequency.\cite{sulewski} This gives the full temperature and
frequency dependent scattering rate, $\gamma^*(\omega,T)\propto
\omega^2+(p \pi T)^2$. Gurzhi calculated $p=2$ for
electron-electron scattering.\cite{gurzhi} For heavy fermion systems,
Sulewski {\it et al.}\cite{sulewski} found that the experimental data
was consistent with $p\leq 1$. They proposed a simple phenomenological
model, $\gamma(\omega)=\gamma_0+\lambda_0 \omega
\omega_0/(\omega-i\omega_0)$, which satisfies the $\omega^2$ behavior
at low frequencies and saturates at a characteristic frequency
$\omega_0$. Using $W$ from the $T^2$ fits to $\gamma^*$ (see
Fig.~\ref{fig;fitparams}) and a value for $p$, $\omega_0$ may be
determined. Taking $p=1$, we find less than a 15\% effect on the mass
enhancement and $\omega_0 \gg 15$~meV, the high-frequency cutoff of
our far-IR measurements ($\omega_0$ values for $p=1$ are
shown in Table~\ref{table1}). Thus, in the far-IR, the frequency
dependence of $\gamma^*$ and $m^*/m_B$ is not significant.

The $T^2$ dependence of $\gamma^*$ and relative $T$-independence of
$\mstar$ differs with results reported on a LCMO polycrystalline
sample.\cite{nohlc} We believe that the discrepancy is due to the
effects of surface damage introduced during the polishing of the bulk
samples.  Recently, Takenaka {\it et al.}\cite{takenaka} and Kim {\it
et al.}\cite {nohns} discussed the importance of polishing and surface
scattering effects in measuring bulk reflectivity for polycrystalline
and single-crystal samples. They found the optical properties of these
materials are sensitive to surface preparation. Specifically,
polishing produced a decrease in mid-IR reflectivity and the apparent
spectral weight. Cleaving or annealing the polished surfaces removes
these effects, bringing the results on bulk samples in reasonable
agreement with thin films.
%The large mid-IR conductivity seen here contrasts with earlier
%results on bulk materials\cite{tokura,nohlc} and is attributed to
%surface preparation problems. 
%Again, we understand the discrepancy with the bulk measurements as
%resulting primarily from the surface problem discussed above.

It is interesting to compare the mass enhancement obtained from
optical measurements reported here with specific heat results. The
low-temperature ($3<T<8$~K) specific heat is $C_H=\gamma_{el} T+\delta
T^{3/2}+\beta T^3$ where the three terms arise from charge carriers,
magnons, and phonons, respectively.\cite{smoly2} The coefficient of
the linear (electronic) term is $\gamma_{el}=\pi^2 k_B^2 N(E_f)/3
\propto m^*$.  For LCMO and LSMO, $\gamma_{el}$ is $4.5 \pm 0.1$ and
$3.4 \pm 0.15$~mJ/mole~K$^2$, respectively. Relating the experimental
$\gamma_{el}$ to band theory predictions gives the specific heat mass
enhancement, $m^*/m_B=\gamma_{el}/\gamma_{el}^B=1+\lambda$, analogous
to the optical enhancement in Eq.~(\ref{eq;firmass}).  Pickett and
Singh\cite{pickett} predict $N(E_f)=0.47$~states/eV for the majority
spin band giving $\gamma_{el}^B=1.1$~mJ/mole~K$^2$. Mass enhancements
calculated using this $\gamma_{el}^B$ are shown in
Table~\ref{table1}. Including the $N(E_f)$ for the minority spin bands
(smaller than majority by approximately a factor of 2) increases
$\gamma_{el}^B$ and $\omega_p^{B\,2}$. This tends to reduce the
specific heat mass enhancement while increasing the optical mass
enhancement. However, minority carriers are expected to be localized
by cation disorder\cite{pickett} and are not included in this report.
A comparison of $m^*/m_B$ from optics ($T=10$~K) and specific heat as
shown in Table~\ref{table1} indicates reasonable agreement between the
two values. This agreement contrasts with the earlier results on
LCMO\cite{nohlc} and LSMO\cite{okimoto,tokura} bulk samples where
large optical masses are reported.  Thus, the agreement we
find between the two mass enhancements suggests that charge ordering
correlations are not strong in the CMR manganite alloys at $x=0.3$
doping.

The strength of the inelastic scattering is large as is indicated by
the small value of $W$ in Table~\ref{table1}: $W \ll E_f \approx
1$~eV. For electron-electron scattering, $W$ is typically on the order
of $E_f$. This is consistent with the relatively large mass
enhancement in these materials in comparison with conventional
metals. The value of the characteristic frequency $\omega_0$ is in
reasonable accord with the expectations of electron-phonon interaction
where $\omega_0$ would be somewhat larger than the average phonon
frequency of the system. Explanations for the temperature dependence
of the scattering rate include magnons\cite{magnons} and
phonons\cite{phonons}. However, the origin of the $T^2$ scattering is
not currently understood.

%\section{Mid-IR}

In the mid-IR, the Fresnel $\begin{mathcal}T\end{mathcal}$ and
$\begin{mathcal}R\end{mathcal}$ formulae for a film on an absorbing
substrate\cite{heavens} are inverted numerically to obtain optical
constants, such as $\sigma(\omega)$ or the dielectric constant
$\epsilon(\omega)$, without the need for Kramers-Kronig
analysis.\cite{wooten} Figure~\ref{fig;mir} shows the real part of the
optical conductivity $\sigma_1$ and the real part of the dielectric
constant $\epsilon_1$ (inset) in the mid-IR for LCMO. The observed
negative $\epsilon_1$ is characteristic of a metal. For $T<150$ K,
$\sigma_1$ exhibits negligible temperature dependence and consequently
we present only the $T=10$ K data (solid line). In this spectral
range, $\sigma_1$ increases at low frequency and eventually must
extrapolate to the far-IR value (solid circle). For comparison, the
Drude conductivity resulting from the fit parameters, $\gamma^*$ and
$\omega_{p}^*$, is shown as a dotted line. The extrapolated
conductivity associated with the Drude model is less than the observed
mid-IR value.

While the mass enhancement analysis does not show an anomalously small
Drude weight, it is interesting to compare the observed far-IR Drude
weight with the total spectral weight from both mid-IR measurements
and band structure calculations. In order to compare these spectral
weights, we introduce the kinetic energy

\begin{equation}  \label{eq;ke}
K(\omega)=\frac{a_0}{e^2}\, S(\omega)=\frac{a_0}{e^2}\ \frac{2}{\pi}
\int_0^\omega \sigma_1(\omega^{\prime})\, d\omega^{\prime} \ ,
\end{equation}

\noindent where $a_0$ is the lattice constant and $S(\omega)$ is the
spectral weight. If $\omega$ is chosen to include all the optical
transitions within the $e_g$ bands $K$ is the kinetic energy of the Mn
$e_g$ electrons. As discussed previously\cite{quijada}, the $e_g$
contribution to the conductivity in these materials is identified to
occur in the frequency range 0~-~2.7~eV.  The measured mid-IR
conductivity for LCMO is combined with data from an earlier sample in
the range 1~-~ 2.7~eV. To estimate $K$, a linear interpolation from
the far-IR data to 0.2 eV is used ($K$ is not sensitive to the choice
of interpolation as the frequency range of the spectral gap is small).
The observed $K$ for LCMO is 280~meV, which agrees with the band
theory prediction, $K_0=0.46\,t_0 \approx 280$~meV.\cite{quijada}
Considering only the contribution from the Drude conductivity,
Eq.~\ref{eq;ke} gives $K_{\text{Drude}}=a_0/(4\pi e^2)\ \wps2$. Thus,
$K_{\text{Drude}}$ from the far-IR is 28~meV, smaller than the band
structure value of 77~meV by the mass enhancement factor $1+\lambda$.
By definition, including frequency dependent scattering in the
extended Drude model to recover the intraband spectral weight leads to
the result shown as a dashed line in Fig.~\ref{fig;mir}. However, as
is clear from Fig.~\ref{fig;mir}, the Drude conductivity alone, even
including the effects of inelastic scattering, is not enough to
account for the large mid-IR conductivity. Only about $1/4$ of the
spectral weight is predicted to be contained in the Drude term. The
remaining spectral weight is most likely coming from interband
transitions occurring between the two Mn $e_g$ bands. The total
infrared conductivity is then the sum of a Drude and interband
contributions, $\sigma=\sigma_{\text{{\tiny
      Drude}}}+\sigma_{\text{{\tiny IB}}}$.

%\section{Conclusions}

We have measured the optical properties of LCMO and LSMO thin films
and obtained $\gamma^*$ and $m^{*}/m_{B}$ as a function of
temperature. $m^{*}/m_{B}$ exhibits little temperature dependence
while $\gamma^*$ shows a strong $T^{2}$ behavior. The optical mass
enhancement differs only slightly from the specific heat mass
enhancement, indicating that charge ordering correlations are not
strong in these materials.  Comparing the Drude weight with the
measured total spectral weight, we found that the infrared
conductivity is in reasonable accord with conventional contributions
from both intraband and interband transitions.

%\section{Acknowledgments}

We thank A. J. Millis for valuable discussions. This work supported in part
by the NSF-MRSEC Grant Nos. DMR-96-32521 and DMR-9705482, and the NSA.

\,

\begin{figure}[tbp]
\caption{Transmittance of LCMO and LSMO as a function of photon energy
for several temperatures. Solid lines are data, dashed lines are fits
to a simple Drude model, and dotted lines are Drude fits in the limit,
${\gamma^*\rightarrow\infty}$. }
\label{fig;data}
\end{figure}

\begin{figure}[tbp]
\caption{Temperature dependence of the scattering rate $\gamma^*$ (a),
plasma frequency $\omega_p^*$ (b), resistivity $\rho$ (c), and mass
enhancement $m^*/m_B$ (d) obtained from fitting the far-IR transmittance
using a Drude conductivity. $\gamma^*$ is fit (solid line)
to $T^2$. The DC resistivity in (c) is plotted for comparison (note for LSMO
the DC resistivity is multiplied by a factor of 9). Error bars are shown
explicitly in (a), (b), and (d) and are nominally the size of the solid
circles in (c). }
\label{fig;fitparams}
\end{figure}

\begin{figure}[tbp]
\caption{Frequency dependence of the real part of the conductivity 
  $\sig1w$ for LCMO. Simple and extended Drude conductivity are shown
  as a dotted and dashed line respectively. Inset shows the real part
  of the dielectric constant $\epsilon_1(\omega)$. }
\label{fig;mir}
\end{figure}

\begin{table}[tbp]
\caption{Mass enhancements from optics ($T=10$~K) and specific heat and 
characteristic scattering-rate energies.}
\label{table1}
\begin{tabular}{lcccc}
& \multicolumn{2}{c}{$m^*/m_B$} & & \\ Sample & Optics &
Specific Heat$^{\text{a}}$ & $W$ (meV) & $\omega_{0}$ (meV) \\
\tableline LCMO & 2.8 $\pm$\ 0.2 & 4.05 $\pm$\ 0.09 & 3.0 & $\leq$ 76
\\ LSMO & 3.0 $\pm$\ 0.3 & 3.06 $\pm$ 0.14 & 3.3 & $\leq$ 87
\end{tabular}
\par
$^{\text{a}}$Ref.~\onlinecite{smoly}
\end{table}

\end{document}